\documentclass[aps,prb,twocolumn,showpacs,superscriptaddress,amssymb,amsmath,floatfix]{revtex4}
\usepackage{graphics,epsfig}

\begin{document}

\title{Dynamic transition and Shapiro step melting in a frustrated Josephson-junction array}

\author{Jong Soo Lim}
\affiliation{Department of Physics, Seoul National University, Seoul 151-747, Korea} 

\author{M.Y. Choi}
\affiliation{Department of Physics, Seoul National University, Seoul 151-747, Korea} 
\affiliation{Korea Institute for Advanced Study, Seoul 130-722, Korea}

\author{J. Choi}
\affiliation{Department of Physics, Keimyung University, Daegu 704-701, Korea} 

\author{Beom Jun Kim}
\affiliation{Department of Molecular Science and Technology, Ajou University,
Suwon 442-749, Korea}

\begin{abstract}
We consider a two-dimensional fully frustrated Josephson-junction array driven 
by combined direct and alternating currents. 
Interplay between the mode locking phenomenon, manifested by giant Shapiro steps 
in the current-voltage characteristics, and
the dynamic phase transition is investigated at finite temperatures.
Melting of Shapiro steps due to thermal fluctuations
is shown to be accompanied by the dynamic phase transition, 
the universality class of which is also discussed.
\end{abstract}

\maketitle

Driven systems with many degrees of freedom are ubiquitous in nature
and exhibit a rich variety of dynamic behavior as the driving or the temperature is varied.  
Here the usual equilibrium concepts are not applicable, making the study of nonequilibrium 
transitions in such driven systems very important and highly nontrivial. 
A prototype system is provided by the Josephson-junction array 
under driving currents,~\cite{Mon,KC,Luo,Marconi,Jeon,Lim} 
which is also closely related to, e.g., moving vortex lattices and 
sliding charge density waves.~\cite{mvl,cdw}
In equilibrium the fully frustrated Josephson-junction array (FFJJA), as an experimental 
realization of the fully frustrated $XY$ model with the U(1)$\times$Z$_2$ symmetry,
has been a source of controversy as to whether a Z$_2$ transition and 
a U(1) transition occur at the same temperature.~\cite{Teitel,ffxy,exponent,Olsson} 
In the presence of direct driving currents, the transition temperatures in general reduce,~\cite{KC} 
approaching zero as the current is increased toward the critical current;
beyond it, the Z$_2$ transition occurs at a substantially lower temperature than 
the U(1) transition.~\cite{Marconi} 
On the other hand, when the driving current has only the ac component,
the FFJJA has been shown to exhibit a dynamic Z$_2$ transition, which belongs to the 
same universality class as the equilibrium Z$_2$ transition.~\cite{Lim} 

The remaining case is the FFJJA driven by combined direct and alternating currents, 
which is well known to display quantized voltage plateaus, called giant Shapiro steps, 
in the current-voltage characteristics at zero temperature.~\cite{Stroud}
This is a manifestation of mode locking in response to the driving current, 
associated with topological invariance of the system.~\cite{Kvale}
However, the issue of the FFJJA driven by {\em combined direct and alternating currents} has not 
been addressed from the viewpoint of the dynamic phase transition, which requires 
analysis of the system {\em at finite temperatures}. 

This work investigates how such a mode locking phenomenon changes into the dynamic
transition, as the temperature is raised from zero.  
To describe the dynamic transition associated with topological invariance, 
we introduce the dynamic order parameter and investigate its behavior with 
the driving amplitude. 
The zero-temperature behavior of the dynamic order parameter is discussed in the framework 
of the topological invariance of Shapiro steps. 
As the temperature is raised from zero, the Shapiro steps are shown to melt, which is
accompanied by the dynamic phase transition. 
From the detailed behavior of the dynamic order parameter at finite temperatures, 
we construct the dynamic phase diagram on the plane of the temperature and the direct current,
and examine nature of the transition.

To begin with, we consider the equations of motion for phase angles \{$\phi_i$\} of 
the superconducting order parameters in grains, which form an $L \times L$ square lattice.
Within the resistively-shunted-junction model 
under the fluctuating twist boundary conditions,~\cite{Kim} they read:
\begin{equation}
{\sum_j}' \left[ \frac{d{\widetilde{\phi}}_{ij}}{dt} + 
\sin({\widetilde{\phi}}_{ij} - {\bf r}_{ij}\cdot{\bf \Delta}) + \eta_{ij} \right] = 0, 
\label{e1} 
\end{equation}
where the primed summation runs over the nearest neighbors of grain $i$ and 
the thermal noise current $\eta_{ij}$ satisfies 
$\langle \eta_{ij}(t) \eta_{kl}(t') \rangle = 
2T\delta(t-t')(\delta_{ik}\delta_{jl} - \delta_{il}\delta_{jk})$ at temperature $T$. 
We have used the abbreviations 
${\widetilde{\phi}}_{ij} \equiv \phi_i - \phi_j - A_{ij}$ and 
${\bf r}_{ij} \equiv {\bf r}_i - {\bf r}_j$
with ${\bf r}_i = (x_i, y_i)$ denoting the position of grain $i$. 
Note that ${\bf r}_{ij}$ for nearest neighboring grains becomes a unit vector 
with the lattice constant set equal to unity. 
We have also written the energy and the time in units of 
$\hbar i_c /2e$ and $\hbar /2eRi_c$, respectively,
with the critical current $i_c$ and the shunt resistance $R$ of a single junction.
The dynamics of the twist variable ${\bf \Delta} \equiv (\Delta_x, \Delta_y)$ 
is governed by the equations  
\begin{align}
& \frac{d\Delta_x}{dt}  
  = \frac{1}{L^2} \sum_{{\langle ij \rangle}_x}\sin({\widetilde{\phi}}_{ij} - \Delta_x) 
    + \eta_{\Delta_x} - I_{dc} - I_{ac}\sin\Omega t \nonumber \\
& \frac{d\Delta_y}{dt}
  = \frac{1}{L^2} \sum_{{\langle ij \rangle}_y}\sin({\widetilde{\phi}}_{ij} - \Delta_y)
    + \eta_{\Delta_y}, 
\label{e2}
\end{align}
where $\sum_{\left\langle ij \right\rangle_\mu}$ denotes the summation over all
nearest neighboring pairs in the $\mu \,(= x, y)$ direction, $\eta_{\Delta_\mu}$ satisfies 
$\langle \eta_{\Delta_\mu}(t)\eta_{\Delta_{\mu'}}(t')\rangle = (2T/L^2)\delta(t-t')\delta_{\mu\mu'}$, 
and the combined direct and alternating currents $I_{dc} + I_{ac} \sin\Omega t$ are injected 
in the $x$ direction. 
In the Landau gauge, the bond angle $A_{ij}$, given by the line integral of the vector potential, 
vanishes for ${\bf r}_j = {\bf r}_i + {\bf \hat{x}}$ and takes the value
$\pi x_i$ for ${\bf r}_j = {\bf r}_i + {\bf \hat{y}}$. 


At zero temperature the FFJJA driven by combined direct and alternating currents
displays integer, fractional, and subharmonic giant Shapiro steps.~\cite{Stroud,Kvale}
The system evolves periodically in time with the topological invariant period, 
characteristic of each voltage step.
Time evolution of the phase configuration shows that the system visits 
periodically various accessible states consisting of
the ground states, the transient states, and other dynamically accessible states 
of the purely dynamic origin.~\cite{Skim} 
To characterize such dynamic behavior associated with the ${\rm Z}_2$ symmetry in the FFJJA,
we consider the chirality
\begin{equation}
C({\bf R}, t) \equiv {\rm sgn} 
\left\{\sum_{\bf P} \sin \left[\widetilde{\phi}_{ij}(t) 
      - {\bf r}_{ij}\cdot {\bf \Delta}(t) \right] \right\} 
\end{equation}
and the staggered magnetization
\begin{equation}
m(t) \equiv \frac{1}{L^2} \sum_{\bf R} (-1)^{x_i + y_i} C({\bf R},t), 
\end{equation}
where $ \sum_{\bf P}$ denotes the directional plaquette summation of links 
around the dual lattice site 
${\bf R} \equiv {\bf r}_i + (1/2)(\hat{\bf x}+\hat{\bf y})$.

\begin{figure}
\epsfig{width=7.3cm,file=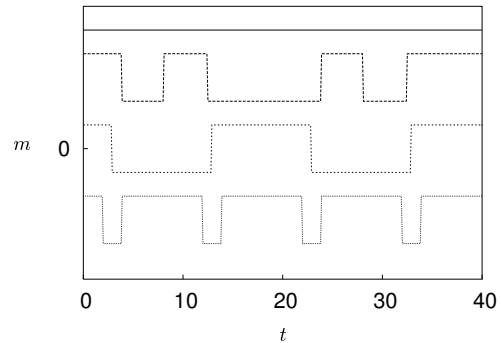} 
\caption [Time evolution of the staggered magnetization $m(t)$ at zero temperature. ]
{Time evolution of the staggered magnetization $m(t)$ at zero temperature.
The amplitude and the frequency of the ac component are $I_{ac} = 1.0$
and $\Omega/2\pi = 0.1$, respectively.
The dc component is given by $I_{dc} = 0.05, 0.177, 0.40$, and $0.75$ from the top to the bottom,
corresponding to the average voltage $\bar{V} = 0, 1/4, 1/2$ and $1$. The period of $m(t)$
is $\tau \,(\equiv 2\pi/\Omega = 10)$, $2\tau$, and $4\tau$ on the steps $\bar{V} = 1, 1/2$ and $1/4$, 
respectively. 
For clarity, data for $\bar{V} = 0$ and $1/4$ have been shifted upward, while those for $\bar{V} = 1$
shifted downward.}
\label{fig1}
\end{figure}

Figure~{\ref{fig1}} exhibits the zero-temperature time evolution of the staggered magnetization
on various voltage steps, 
where $\bar{V}$ denotes the time-averaged voltage in units of $L\Omega \hbar /2e$.
The period of $m(t)$ in Fig.~\ref{fig1} is given by $\tau$, $2\tau$, and $4\tau$ 
on steps $\bar{V}$ = 1, 1/2, and 1/4, respectively.
Here we have considered the FFJJA of size up to $L=24$, and integrated numerically 
the equations of motion in Eqs.~(\ref{e1}) and (\ref{e2}) via the modified Euler method, 
using time steps of size $\Delta t = 0.01$.  The values of $\Delta t$ have been varied, 
only to give insignificant difference. 
Typically, data have been averaged over 1000 driving periods of the (alternating) current, after
the initial 500 periods discarded; 
this has turned out sufficient for reaching appropriate stationarity.

At high temperatures thermal fluctuations are so strong that
the influence of the driving current and the lattice potential can be neglected,
leading to random fluctuations of $m(t)$. 
We thus expect a phase transition between the dynamically ordered phase
and the disordered one as the temperature is varied. 
To describe such a dynamic phase transition, we introduce the dynamic order parameter, 
defined to be the staggered magnetization averaged over $n$ periods of the (alternating) current:
\begin{equation}
\label{eq:Q}
Q \equiv \frac{\Omega}{2\pi n} \left| \int_{t_0}^{t_0+n \tau } m(t) \,dt \right|
\end{equation}
with $\tau \equiv 2\pi / \Omega$. 
On each Shapiro step, we consider the topologically invariant period~\cite{Skim}
to choose the proper integer $n$.  Specifically, on the $1/2q$ step we choose $n = q$,
avoiding the null order parameter.  In general, the dynamic order parameter in Eq.~(\ref{eq:Q}) 
depends on $t_0$ and henceforth the maximum value is defined to be the dynamic order parameter:
$Q \equiv \max_{t_0} Q(t_0)$. 

\begin{figure}
\epsfig{width=7.7cm,file=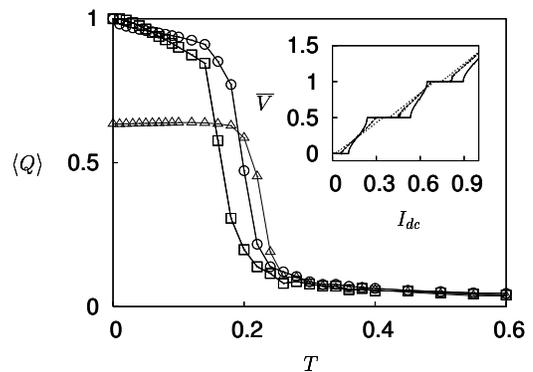} 
\caption[Dynamic order parameter as a function of the temperature]
{Dynamic order parameter as a function of the temperature in the system of size $L = 24$
for $I_{dc} = 0.04(\Box), 0.4(\bigcirc)$, and $0.75(\triangle)$, corresponding
to $\bar{V} = 0, 1/2,$ and $1$, respectively.  Lines are merely
guides to eyes.  Inset: Time averaged voltage $\bar{V}$ versus the dc
component $I_{dc}$ at temperature $T=0, 0.1$, and 0.24.  As $T$ is raised, 
the voltage plateaus tend to melt.}
\label{fig2}
\end{figure}
Figure~\ref{fig2} shows the ensemble average $\langle Q\rangle$ of the dynamic 
order parameter as a function of the temperature $T$ at various values of the dc component $I_{dc}$. 
Manifested is the presence of a dynamic phase transition, separating the
dynamically ordered phase at low temperatures and the disordered phase at high temperatures.
In all the three cases in Fig.~\ref{fig2}, corresponding to the voltage steps 
$\bar{V} = 0, 1/2,$ and $1$, the dynamic order parameter increases gradually
as the temperature is lowered from high temperatures. 
Subsequently, it grows rapidly, eventually saturating to the zero-temperature value. 
Further, it is also observed that cooling and heating curves for the dynamic order
parameter exhibit appreciable hysteresis at low temperatures (not shown here), 
which arises from the asymmetry of the lattice potential induced by the dc component
of the driving current and the resulting anisotropy, similarly to the dc driven case.~\cite{Marconi} 
Such behavior does not appear in the system driven by alternating currents only (without the dc component). 

\begin{figure}
\centering
\epsfig{width=7.7cm,file=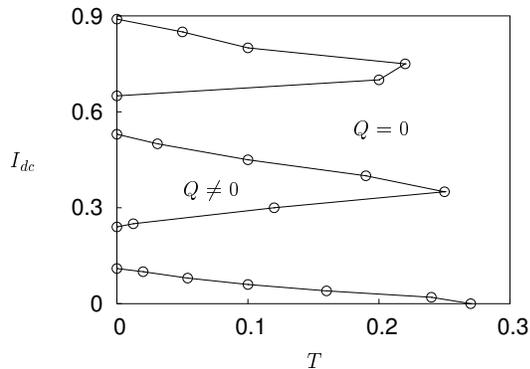}
\caption[Dynamic phase diagram]{Dynamic phase diagram on the $T{-}I_{dc}$ plane.
The phase boundaries are determined by the crossing points of Binder's cumulants
for size $L = 8, 16$, and $24$.
The areas enclosed by the boundaries specify dynamically ordered phases characterized by $Q\neq 0$.  
Lines are merely guides to eyes.}
\label{fig3}
\end{figure}
The Shapiro step originates from the interplay between the period of the external driving current 
and the period emerging from the intrinsic dynamics of the system. 
Such mode locking behavior present in the low temperature regime is 
expected to smear out as the thermal fluctuations become strong. 
Indeed the Shapiro steps are observed to melt as the temperature is raised, 
disappearing eventually at sufficiently high temperatures
(see the inset of Fig.~\ref{fig2}). 
To probe the relation between this melting behavior and the dynamic phase transition, 
we first estimate the transition temperature accurately via Binder's cumulant~\cite{Binder}
\begin{equation}
U_L = 1 - \frac{\langle Q^4 \rangle}{3{\langle Q^2 \rangle}^2},
\end{equation}
which takes the value  $2/3$ at $T=0$, vanishes in the high-temperature limit,
and becomes size-independent at the transition temperature.  
The unique crossing point of $U_L$, independent of the system size $L$, 
for given $I_{dc}$ thus yields the corresponding data point for the phase
boundary on the $T{-}I_{dc}$ plane in Fig.~\ref{fig3}.  
The areas enclosed by the resulting phase boundaries specify the dynamically ordered phase, 
characterized by $Q \neq 0$, for $\bar{V} = 0$, $1/2$, and $1$ from the bottom to the top.
The transition temperature initially decreases monotonically to zero as the
driving amplitude $I_{dc}$ is increased in the region of $\bar{V} =
0$.  Further increase of $I_{dc}$ drives the system into another ordered
region, characterized by $\bar{V} = 1/2$, where the transition
temperature first grows with the driving current, then reduces to zero, and so on.

\begin{figure}
\centering
\epsfig{width=7.7cm,file=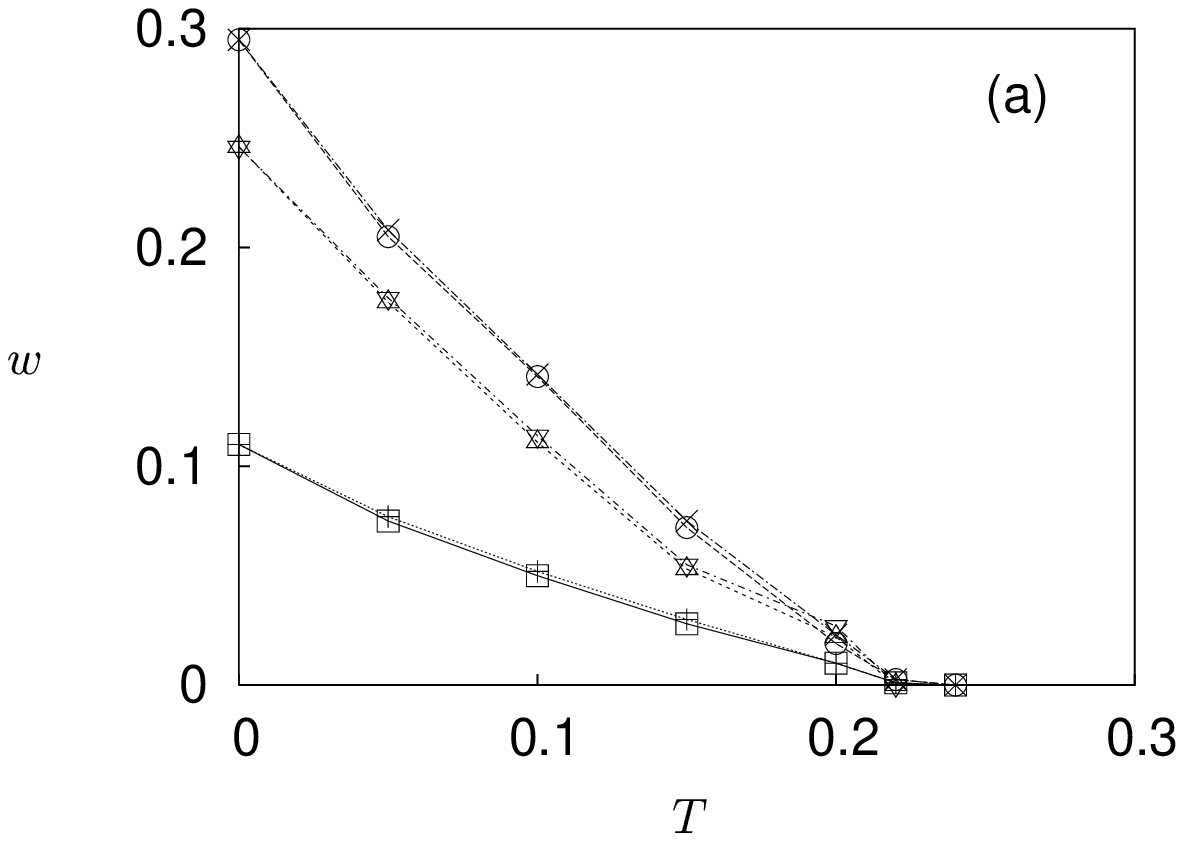}
\epsfig{width=7.7cm,file=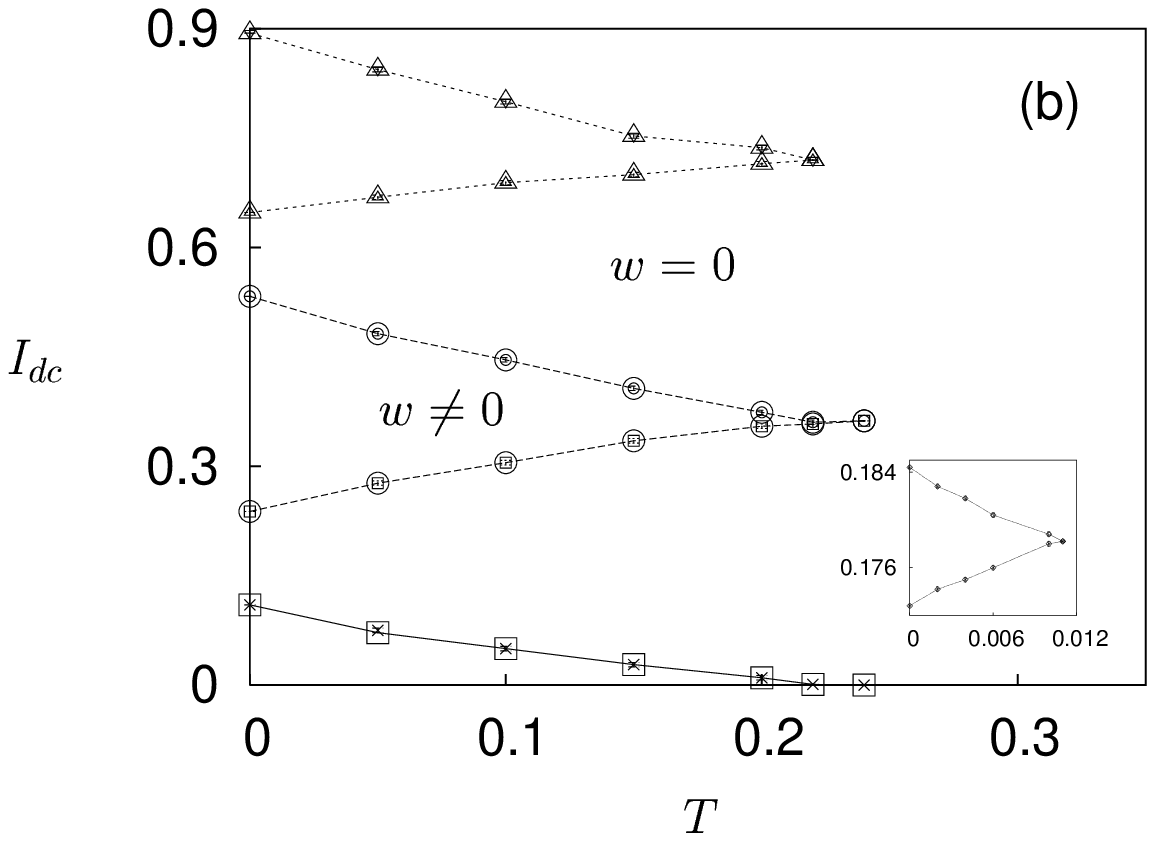}
\caption[Dynamic phase diagram]{
(a) Width $w$ of the voltage plateau versus the temperature $T$ on the step 
$\bar{V} = 0(\square,+), 1/2(\bigcirc,\times)$, and $1(\triangle,\triangledown)$,
where the two symbols for each step represent the data from two different criteria of the slope: 
$0.005$ and $0.01$, respectively. 
(b) Dynamic phase diagram on the $T{-}I_{dc}$ plane, 
determined by the width of voltage plateaus for size $L = 16$.
The areas enclosed by the phase boundaries specify dynamically ordered phases with $w \neq 0$.
Lines are merely guides to eyes.  Inset: Detailed diagram displaying the ordered phase
corresponding to the subharmonic voltage step $\bar{V} =1/4$, again determined by its width.}
\label{fig4}
\end{figure}
For comparison, phase boundaries are also estimated independently from melting of the Shapiro steps. 
Here we define the width of a step as follows:
For a given point on a voltage step, we measure the slope of the straight line
connecting the point to the middle point of the step and consider the point to be
located on the step if the slope is less than 0.005.  
Namely, the step width is determined by the boundary point, for which the slope is given by 0.005
[the use of other values hardly changes the resulting phase boundaries, 
as shown in Fig.~\ref{fig4}(a)]. 
The step width obtained in this manner diminishes with the temperature and eventually
vanishes to zero [see Fig.~\ref{fig4}(a)], from which 
each phase boundary in Fig.~\ref{fig4}(b) has been estimated. 
The good agreement between Figs.~\ref{fig3} and \ref{fig4}(b)
manifests the correspondence between the nonzero value of the dynamic order parameter 
and the finite width of the Shapiro step, thus
indicating that the dynamic phase transition accompanies melting of Shapiro steps. 
It is also expected that there exist additional dynamically ordered phases,
associated with the series of subharmonic Shapiro steps;~\cite{Skim} 
they presumably form Arnold tongue structure~\cite{circle} in the complete phase diagram. 
We have thus examined the subharmonic step $\bar{V} = 1/4$, and indeed found the 
corresponding ordered phase, shown in the inset of Fig.~\ref{fig4}(b). 
Unfortunately, other phases, corresponding to higher-order subharmonic steps, 
occupy even tinier regions, making it very difficult to locate the boundaries. 

\begin{figure}
\centering
\epsfig{width=7.7cm,file=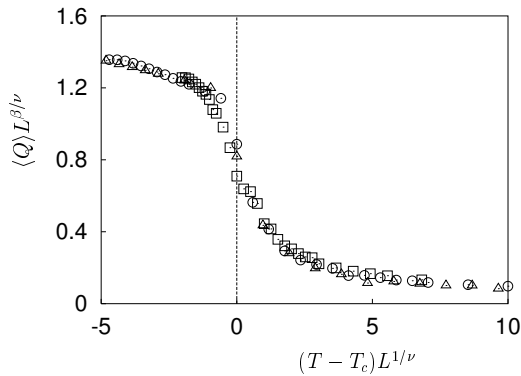} 
\caption[Scaling plot of the dynamic parameter]
{Scaling plot of the dynamic order parameter for size $L = 8(\Box),\,16(\bigcirc)$, and $24(\triangle)$
at dc component $I_{dc} = 0.04$. 
The fitting has been performed with the critical exponents $\nu = 0.82$ and $\beta/\nu = 0.11$.}
\label{fig5}
\end{figure}

We finally study the nature of the transition by means of finite-size scaling for 
the dynamic order parameter:
\begin{equation}
\langle Q \rangle = L^{-\beta/\nu} f((T{-}T_c)L^{1/\nu}) .
\end{equation}
In Fig.~\ref{fig5} we plot $\langle Q \rangle L^{\beta/\nu}$ versus $(T -T_c)L^{1/\nu}$ 
at $I_{dc} = 0.04$ (corresponding to $\bar V =0$), 
which yields reasonable scaling collapse with the critical exponents 
$\nu = 0.82$ and $\beta/\nu = 0.11$.
Since these values of the critical exponents agree well with those for the equilibrium ${\rm Z}_2$ 
transition in the FF$XY$ model,~\cite{Luo,exponent} we conclude that the dynamic phase transition 
in the system here belongs to the same universality class as the equilibrium ${\rm Z}_2$ transition.
Similar conclusion was also reached in the FFJJA under weak staggered oscillating fields 
or uniform alternating currents.~\cite{Jeon,Lim} 
This suggests that the universality class of the transition does not change 
if $\bar V$ remains to be zero (i.e., in the same Arnold tongue). 
On the other hand, at larger values of $I_{dc}$ corresponding to $\bar V \ne 0$, 
the scaling plot does not collapse well and there still exists ambiguity 
on the nature of the transition; 
this appears similar to the dc driven case, where the transition nature changes as the driving 
current exceeds the critical value.~\cite{KC,Marconi}

In summary, we have examined the 2D FFJJA driven by combined direct and alternating currents, as 
a prototype example of driven systems exhibiting rich dynamic behaviors. 
The relation between melting of the voltage plateaus (Shapiro steps) and the dynamic phase transition 
has been explored. 
It has been observed that melting of the Shapiro steps is accompanied by the dynamic phase transition,
revealing the correspondence between the two phenomena. 
We have also examined the nature of the transition, to find, on the zero voltage step, 
the same universality class as the equilibrium Z$_2$ transition.

This work was supported in part by 
KOSEF grants R01-2002-000-00285-0 and R14-2002-062-01000-0 as well as by
the BK21 Program.


\end{document}